\documentclass[prb,aps,showpacs,twocolumn,nobibnotes,epsf]{revtex4}

\usepackage{graphicx}
\usepackage{dcolumn}
\usepackage{bm}
\usepackage{SIunits}

\begin{document}

\title{Superconductivity in alkali-earth metals doped phenanthrene}
\author{X. F. Wang$^1$, Y. J. Yan$^1$, Z. Gui$^2$, R. H. Liu$^1$, J. J. Ying$^1$, X. G. Luo$^1$ and X. H. Chen$^1$}
\altaffiliation{Corresponding author} \email{chenxh@ustc.edu.cn}
\affiliation{1.Hefei National Laboratory for Physical Science at
Microscale and Department of Physics, University of Science and
Technology of China, Hefei, Anhui 230026, People's Republic of
China\\
2. State Key Laboratory of Fire Science, University of Science and
Technology of China, Hefei, Anhui 230026, China}

\begin{abstract}
We discover superconductivity in alkali-earth metals doped
phenanthrene. The superconducting critical temperatures \emph{T}$_c$
are 5.6 K and 5.4 K for Sr$_{1.5}$phenanthrene and
Ba$_{1.5}$phenanthrene, respectively. The shielding fraction of
Ba$_{1.5}$phenanthrene exceeds 65\%. The Raman spectra show 8
cm$^{-1}$/electron and 7 cm$^{-1}$/electron downshifts for the mode
at 1441 cm$^{-1}$ due to the charge transfer to organic molecules
from the dopants of Ba and Sr. Similar behavior has been observed in
A$_3$phenanthrene and A$_3$C$_{60}$(A = K and Rb). The positive
pressure effect in Sr$_{1.5}$phenanthrene and Ba$_{1.5}$phenanthrene
together with the lower $T_c$ with larger lattice indicates
unconventional superconductivity in this organic system.

\end{abstract}
\vskip 300 pt

\maketitle

Superconductors, materials that conduct electricity without
resistance and completely eject the magnetic field
lines\cite{handbook}, are mostly inorganic
materials\cite{onnes,nd3sn,nd3ge,cuprate,iron}. Organic
superconductors are very intriguing in condensed matter physics
community due to low dimensionality, strong electron-electron and
electron-phonon interactions and the proximity of
antiferromagnetism, insulator states and superconductivity.
Basically, there are mainly two types of organic superconductors:
i): the quasi-one-dimensional Bechgaard and Fabre salts-
(TMTSF)$_2$X (TMTSF = tetramethyl tetraselena
fulvalene)\cite{Jerome} and (TMTTF)$_2$X (TMTTF =
tetramethyltetrathiafulvalene, X = monovalent anions)\cite{Parkin};
ii): quasi-two-dimensional salts derived from the donor molecule
(BEDT-TTF)$_2$X (BEDT-TTF =
[bis(ethylenedithio)tetrathiafulvalene])\cite{Kini}. The recent
discovery of superconductivity in doped organic crystals containing
an extended phenanthrene-like structural motif\cite{Mallory}, which
is designated as [n]phenacens(n = 3\cite{xfwang} and
5\cite{SC-picene}), has provided new empirical substance to the
occurrence of superconductivity in organic, $\pi$-molecular
materials. Organic materials are generally considered as electrical
insulators. Both phenanthrene (n = 3) and picene (n = 5) are
semiconductors with band gaps of 3.16 eV\cite{Tariq BHATT} and 3.3
eV\cite{SC-picene}, respectively. Superconductivity is introduced by
doping alkali-metals into the interstitial sites of the pristine
compounds. The charge (electron) transfer from alkali-metal atoms to
the molecules results in changes of the electronic structure and the
physical properties to realize superconductivity.

All the reported organic superconductors, including the doped
phenanthrene type, contain five-member rings or six member rings
with conjugated $\pi$-orbital interactions among these rings. The
$\pi$-electron can delocalize throughout the crystal, giving rise to
metallic conductivity due to a $\pi$-orbital overlap between
adjacent molecules. The molecule and crystal structure of organic
materials are completely different from the inorganic system, which
makes the understanding of superconductivity in organic system quite
difficult and complicated. The superconductivity in doped
phenanthrene-type materials offers an excellent candidate for
studying physics in organic, $\pi$-molecular superconductors.
However, the superconducting fraction was reported to be rather
small in both phenanthrene and picene system. The maximum shielding
fractions in the powder samples of  K-doped picene and phenanthrene
are 1.2\%~\cite{SC-picene} and 5.3\%~\cite{xfwang}, respectively. This
makes it difficult to investigate intrinsic superconducting
properties in phenanthrene-type systems. Obtaining high-quality
samples and pursuing new superconductors in this hydrocarbon
superconducting family are the central issues for investigating
intrinsic physical properties and understanding the superconducting
mechanism in organic, $\pi$-molecular superconductors. Here we
report the discovery of superconductivity in strontium and barium
doped phenanthrene, and the superconducting transition temperature
is \emph{T}$_c$ of 5.6 K and 5.4 K for Sr$_{1.5}$phenanthrene and
Ba$_{1.5}$phenanthrene, respectively. The shielding fraction is up
to 65.4\% in Ba$_{1.5}$phenanthrene at 2 K. Raman spectra show 8
cm$^{-1}$/electron and 7 cm$^{-1}$/electron downshifts due to the
charge transfer, which are similar to those of $A_3$phenanthrene and
$A_3$C$_{60}$($A$=K and Rb). The pressure dependence of
superconductivity shows a positive coefficient
\emph{d(T}$_{C}$\emph{/T}$_{C}$(0)\emph{)/dP} and an enhancement of
the shielding fraction.

\begin{figure}[h]
\centering
\includegraphics[width=0.5\textwidth]{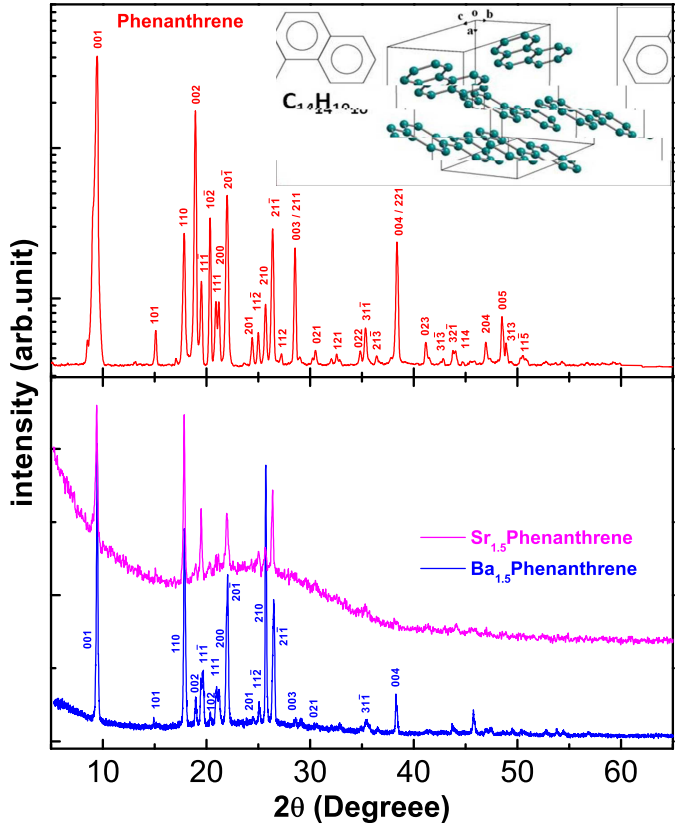}
\caption{X-Ray diffraction patterns for phenanthrene,
Sr$_{1.5}$phenanthrene and Ba$_{1.5}$phenanthrene. (a): X-ray
diffraction patten for pristine phenanthrene, the molecule and
crystal structure are shown in the inset. (b): X-ray diffraction
patterns for Sr- and Ba-doped phenanthrene.} \label{fig1}
\end{figure}

Barium (99\%), Strontium (99\%), phenanthrene (98\%) were purchased
from Alfa Aesar. The phenanthrene was purified by sublimation method
\cite{xfwang}. Barium and strontium were ground into powder with
file. The purified phenanthrene and Ba/Sr powder were mixed with
chemical stoichiometry ratio. The mixture was ground carefully and
then pressed into pellets. The samples were sealed in quartz tube
under vacuum less than 5$\times$10$^{-4}$ Pa. The sample was heated
at 230$\celsius$ for 8 days with intermediate grinding and pelleting
for three times. Finally, the products with uniform dark black color
were obtained. The X-ray diffraction and Raman measurement were
carried out by sealing the samples in capillaries made of special glass No. 10 and purchased from Hilgenberg GmbH.

\begin{figure}[h]
\centering
\includegraphics[width=0.5\textwidth]{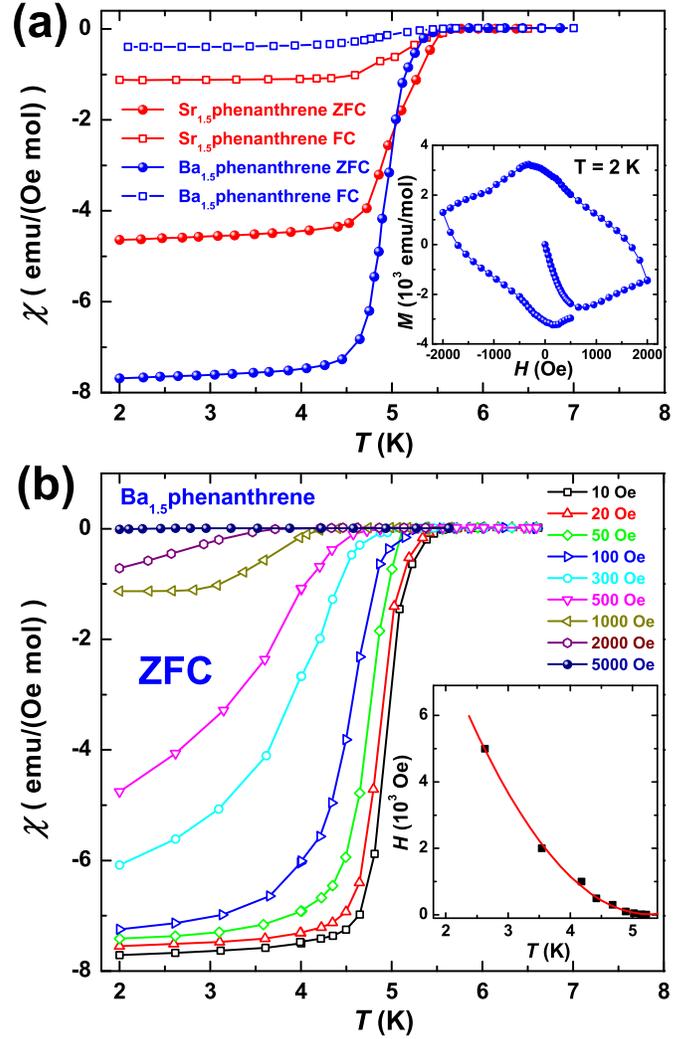}
\caption{Temperature dependence of magnetic susceptibility ($\chi$) for
Sr$_{1.5}$phenanthrene and Ba$_{1.5}$phenanthrene. (a): $\chi$
versus T plots for Sr$_{1.5}$phenanthrene and Ba$_{1.5}$phenanthrene
in the zero-field-cooling (ZFC) and field-cooling (FC) measurements
under the magnetic field of 10 Oe. The inset shows M-H curve of
Ba$_{1.5}$phenanthrene at 2 K. (b): $\chi$ versus \emph{T} curve for
Ba$_{1.5}$phenanthrene in the ZFC measurements under different
magnetic fields. The H versus T$_C$ plot is shown in the inset.}
\label{fig2}
\end{figure}

\begin{figure}[h]
\centering
\includegraphics[width=0.4\textwidth]{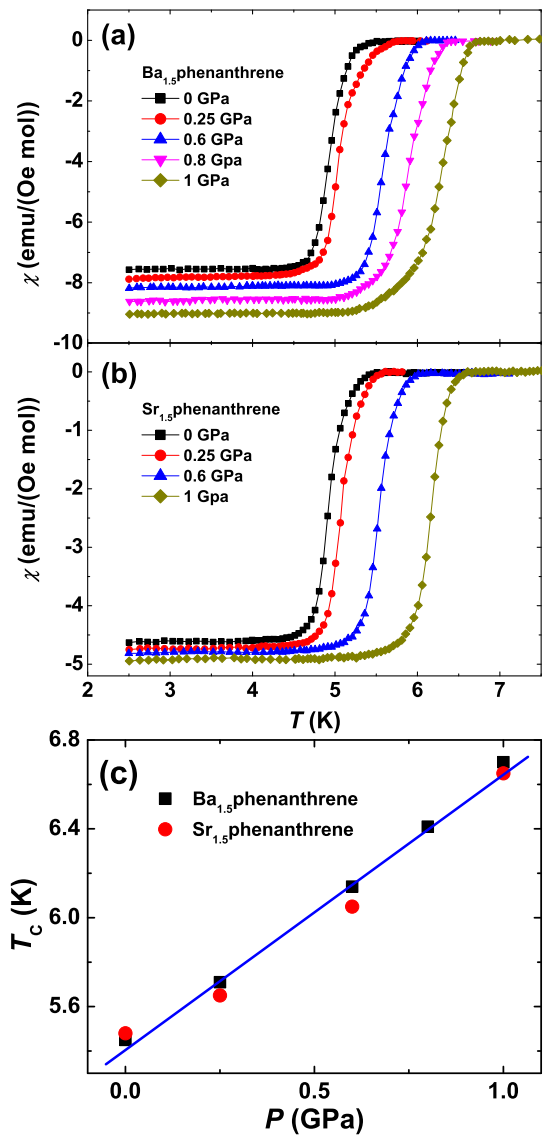}
\caption{Pressure dependence of superconducting transition temperature
\emph{T}$_{C}$ for Sr$_{1.5}$phenanthrene and
Ba$_{1.5}$phenanthrene. (a). Magnetic susceptibility $\chi$ versus \emph{T} in
ZFC measurements for Ba$_{1.5}$phenanthrene under the pressures of P
= 0, 0.25, 0.6, 0.8 and 1.0 GPa. (b). Magnetic susceptibility $\chi$ versus T
in ZFC measurement for Sr$_{1.5}$phenanthrene under different
pressures of p = 0, 0.25, 0.6 and 1.0 GPa. (c). Pressure dependence
of \emph{T}$_C$ for superconducting Ba$_{1.5}$phenanthrene and
Sr$_{1.5}$phenanthrene.} \label{fig3}
\end{figure}

Figure 1 shows the X-ray diffraction (XRD) patterns of the pure, and
Ba- and Sr- doped phenanthrene, respectively. Fig.1(a) shows the XRD
pattern of pure phenanthrene. There are three fused benzene rings in
the molecule of phenanthrene, as shown in the inset of Fig.1(a). The
phenanthrene crystallizes in the space group of
\emph{P}$_2$$_{\emph{1}}$. The lattice parameters for pristine
phenanthrene are a = 8.453${\rm \AA}$, b = 6.175${\rm \AA}$, c = 9.477${\rm \AA}$ and
$\beta$=98.28$^{\degree}$, being consistent with the results reported
before\cite{trotter}. Fig.1(b) shows the XRD patterns of Sr- and Ba-doped phenanthrene, respectively. All the peaks can be well indexed
with the \emph{P}$_2$$_{\emph{1}}$ space group. No impurity phase
was found in the XRD pattern. Lattice parameters are a = 8.471
${\rm \AA}$, b = 6.181${\rm \AA}$, c = 9.491${\rm \AA}$, $\beta$ = 97.55$^{\degree}$ for
Sr$_{1.5}$phenanthrene; a = 8.479${\rm \AA}$, b = 6.177${\rm \AA}$, c =
9.502${\rm \AA}$, $\beta$ = 97.49$^\degree$ for Ba$_{1.5}$phenanthrene. It
indicates that the superconducting phase discussed below is
responsible for the superconductivity. The lattice parameters show
slightly changes compared to pristine phenanthrene. The unit cell
volume increases slightly from 489.5 ${\rm \AA}^3$ for phenanthrene to
492.6 ${\rm \AA}^3$ for Sr$_{1.5}$phenanthrene and 493.4 ${\rm \AA}^3$ for
Ba$_{1.5}$phenanthrene, which is consistent with doped TMTSF case.
With monovalent anions injection, the unit cell volume of
(TMTSF)$_2$PF$_6$ expands to 345.5 {\rm \AA}$^3$ per TMTSF molecule from
316 {\rm \AA}$^3$ per TMTSF in the pristine compound\cite{Kim}. While in
K$_x$picene, the unit cell volume shows significantly shrinkage due
to potassium doping\cite{SC-picene}. Generally, there are two
effects of foreign atom/ionic intercalation on unit cell volume.
Firstly, the lattice will expand because of the volume of foreign
atom. Secondly, the attracting force becomes electrostatic
attraction from Van der Waals force, which will reduce the unit cell
volume. The change of lattice parameter induced by intercalation of
Sr and Ba indicates former factor contributes more in Sr- and Ba-doped phenanthrene to affect the crystal structure.

\begin{figure}[h]
\centering
\includegraphics[width=0.4\textwidth]{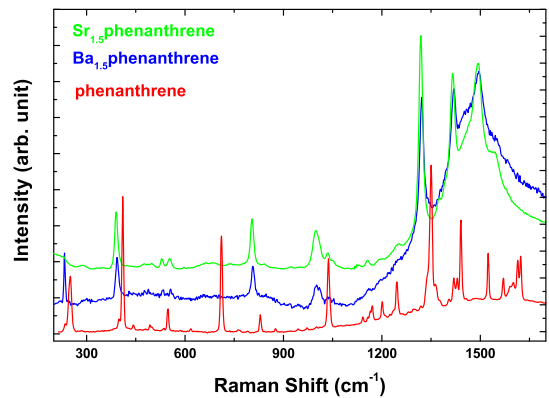}
\caption{Raman spectra for the pure phenanthrene, the
superconducting Sr$_{1.5}$phenanthrene and Ba$_{1.5}$phenanthrene. }
\label{fig4}
\end{figure}

Figure 2(a) shows the temperature dependence of magnetic
susceptibility $\chi$ for the powder samples of Sr$_{1.5}$phenanthrene
and Ba$_{1.5}$phenanthrene in the zero-field cooling (ZFC) and
field-cooling (FC) processes under the magnetic field of 10 Oe.
Susceptibility shows a drastic drop in ZFC and FC measurements at
5.6 K and 5.4 K for Sr$_{1.5}$phenanthrene and Ba$_{1.5}$phenanthrene,
respectively. The temperature corresponding to the sharp drop is
defined as the superconducting transition temperature ($T_c$).  The
superconducting transition temperatures are higher than the $T_c$ of
4.95 K and 4.75 K for K$_3$phenanthrene and Rb$_3$phenanthrene,
respectively. The intercalation of larger atom results in a lower
\emph{T}$_C$, which is consistent with the alkali-metal doped
phenanthrene. The evolution of \emph{T}$_C$ with the radii of
intercalating atoms for doped phenanthrene is quite different from
the doped C$_{60}$, whose superconducting transition temperatures
increase monotonically as the unit cell size
increases\cite{fleming}. It cannot be understood in BCS theory
because the density of states (DOS) increases with lattice
expansion, and increase of DOS should lead to enhancement of $T_C$
based on BCS theory. However, $T_C$ of Sr$_{1.5}$phenanthrene is
little higher than that of Ba$_{1.5}$phenanthrene with larger unit
cell volume. In this sense, Sr$_{1.5}$phenanthrene and
Ba$_{1.5}$phenanthrene superconductors should be unconventional.
Diamagnetic signals from zero-field-cooling and field-cooling
measurements can be assigned to shielding and Meissner effects. As
shown in Fig.2(a), the shielding fraction and the Meissner fraction
are 65.4\% and 3.4\% for the powder sample of
Ba$_{1.5}$phenanthrene, 39.5\% and 9.5\% for the sample of
Sr$_{1.5}$phenanthrene, respectively. The shielding fraction is much
larger than that of alkali-metal doped picene and phenanthrene. The
superconducting transition width is $\sim$ 0.8 K, which is comparable
with K$_{3}$phenanthrene and Rb$_{3}$phenanthrene. The \emph{M}
versus \emph{H} plot is shown as the inset of Fig.2(a). The lower
critical field \emph{H}$_{C1}$ of Ba$_{1.5}$phenanthrene is 720 Oe
at 2 K, which is larger than the value of 380 Oe at 5 K in
K$_{3.3}$picene with \emph{T}$_{C}$ of 18 K and 175 Oe at 2 K in
K$_{3}$phenanthrene with \emph{T}$_{C}$ of 5 K .

Figure 2(b) shows $\chi$ versus \emph{T} plots for
Ba$_{1.5}$phenanthrene superconductor under different magnetic
fields. The diamagnetic signal is gradually suppressed and the
superconducting transition became significantly broad with the
application of magnetic fields. One can clearly observe a
superconducting transition at 3.5 K with the magnetic field up to
2000 Oe. When the field is higher than 5000 Oe, it is difficult to
observe the superconducting transition from magnetization curve. The
magnetic field H versus \emph{T}$_C$ plot is shown in the inset of
Fig.2(b). It is difficult to precisely determine the upper critical
field \emph{H}$_{C2}$ from \emph{H-T$_{C}$} curve, but it is obvious
that it exceeds 2000 Oe.

Figure 3(a) and 3(b) show the temperature dependence of $\chi$ for
Sr$_{1.5}$phenanthrene and Ba$_{1.5}$phenanthrene in ZFC
measurements under different pressures, respectively. The most
remarkable result is that \emph{T}$_{C}$ for both of
Sr$_{1.5}$phenanthrene and Ba$_{1.5}$phenanthrene increases with
increasing applied pressure. The shielding fraction also becomes
larger. It indicates that the superconductivity is enhanced with the
pressure. As shown in Fig.3(c),
\emph{d(T}$_{C}(P)$\emph{/T}$_C(0)$\emph{)/dP} are $\sim$0.21
GPa$^{-1}$ and 0.23 GPa$^{-1}$ for Sr$_{1.5}$phenanthrene and
Ba$_{1.5}$phenanthrene, respectively. The positive pressure effect
is similar to that of K$_{3}$phenanthrene. According to BCS theory,
the decrease of \emph{T}$_{C}$ is expected with application of
pressure because application of pressure leads to contraction of
lattice, and consequently  density of states \emph{N(E$_{F}$)} at
the Fermi level decreases. The unusual positive pressure effect
indicates unconventional superconductivity in Sr$_{1.5}$phenanthrene
and Ba$_{1.5}$phenanthrene superconductors.

Figure 4 shows the Raman scattering for pristine phenanthrene,
Sr$_{1.5}$phenanthrene and Ba$_{1.5}$phenanthrene. There are seven
major peaks in pristine phenanthrene: 1524, 1441, 1350, 1037, 830,
411 and 250 cm$^{-1}$. The major peaks of the Raman spectrum belong
to A$_{1}$ mode due to the C-C stretching
vibration\cite{Martin,Godec}. The phonon modes in
Sr$_{1.5}$phenanthrene and Ba$_{1.5}$phenanthrene show redshift
relative to the pristine phenanthrene. Such phonon-mode softening
effect arises from the charge transfer into phenanthrene molecule
from dopants of Ba and Sr. The main mode shifts from 1441 cm$^{-1}$
in pristine phenanthrene to 1416 cm$^{-1}$ in Sr$_{1.5}$phenanthrene
and 1419 cm$^{-1}$ in Ba$_{1.5}$phenanthrene. There are 25 and 22
cm$^{-1}$ downshifts for Sr$_{1.5}$phenanthrene and
Ba$_{1.5}$phenanthrene, respectively. It indicates 8
cm$^{-1}$/electron  downshift for Sr$_{1.5}$phenanthrene and 7
cm$^{-1}$/electron downshift for Ba$_{1.5}$phenanthrene due to the
charge transfer. This is consistent with that of K$_3$phenanthrene(6
cm$^{-1}$/electron) and Rb$_3$phenanthrene(7 cm$^{-1}$/electron).
Such similar behavior has been observed in $A_{3}$C$_{60}$ ($A$=K, Rb,
Cs), in which 6 cm$^{-1}$/electron redshift at 1460 cm$^{-1}$
occurs\cite{zhou}. The Raman shift induced by charge transfer is
nearly the same, and independent of dopants, being similar to that
of alkali-metal doped C$_{60}$.

In summary, we discover the superconductivity in Sr- and Ba-doped
phenanthrene-type system. Sr$_{1.5}$phenanthrene and
Ba$_{1.5}$phenanthrene show superconductivity at 5.6 K and 5.4 K,
respectively. The positive pressure effect in Sr$_{1.5}$phenanthrene
and Ba$_{1.5}$phenanthrene and lower $T_c$ with larger lattice
indicate unconventional superconductivity in this organic system.
The superconductivity can be realized by doping alkali-earth metals
to provide charge  and transfer it to organic molecules.

\end{document}